\documentclass[aps, twocolumn,showpacs,superscriptaddress,groupedaddress,floatfix]{revtex4-1}  % for review and submission
\usepackage{graphicx}  % needed for figures
\usepackage{dcolumn}   % needed for some tables
\usepackage{bm}        % for math
\usepackage{amssymb}
\usepackage{amsmath}   % for math
\usepackage{hyperref}
\usepackage[dvipsnames]{xcolor}
\pdfoutput=1
\hypersetup{colorlinks,linkcolor={blue},citecolor={blue},urlcolor={blue}}
\newcommand{\ns}{n_{\rm s}}

\newcommand{\ie}{\textit{i.e.,~}}

\newcommand{\tV}{\tilde{V}}
\newcommand{\tphi}{\tilde{\phi}}
\newcommand{\mpl}{M_{\rm pl}}

\begin{document}
\title{Einstein or Jordan: seeking answers from the reheating constraints}
\author{Debottam Nandi}
\email{E-mail: debottam@physics.iitm.ac.in}
\author{Pankaj Saha}
\email{E-mail: pankaj@physics.iitm.ac.in}
\affiliation{Department of Physics, Indian Institute of Technology Madras, 
Chennai 600036, India}
%%%%%%%%%%%%%%%%%%%%%%%%%%%%%%%%%%%%%%%%%%%%%%%%%%%%%%%%%%%%%%%%%%%%%%%%%%%%%%%
\begin{abstract}
Distinguishing conformally coupled frames from the tree-level perturbative observables (scalar spectral index $\ns$ and tensor-to-scalar ratio $r$) is challenging in cosmology as they are nearly identical. However, since the background evolution in these two frames differs significantly, we can look for potential signatures in the reheating constraints to discriminate these frames. In this work, we study the reheating phase in these frames and find that the difference in the inflationary energy scales in these frames contributes to a significant difference in the reheating e-folding number and hence, different reheating temperature. This difference will eventually lead to a contrasting thermal history in the two frames, which may have a potential observational signature in future observations. This study will open up an avenue for distinguishing various conformally connected otherwise indistinguishable frames and may finally lead us to the correct theory of gravity for our Universe.
\end{abstract}
\maketitle

%%%%%%%%%%%%%%%%%%%%%%%%%%%%%%%%%%%%%%%%%%%%%%%%%%%%%%%%%%%%%%%%%%%%%%%%%%%%%%%

\par{\it Introduction:} Einstein's general theory of relativity provides a compelling and testable theory of the evolution of our universe: the standard model of cosmology, also otherwise known as the {\it Big-Bang} theory. Till date, it has withstood all the experimental tests \cite{Aghanim:2018eyx, *Abbott:2018lct}. Also, its extension to the early Universe, \ie the inflationary paradigm are in good agreement with the present experiments with unprecedented precision \cite{Akrami:2018odb}.
%\cite{STAROBINSKY198099, *STAROBINSKY1982175, Guth:1981, *Guth:1982, *Sato:1981, *LINDE1982389, *Linde:1983gd, *Albrecht-Steinhardt:1982, *Mukhanov:1981xt, *HAWKING1982295, *VILENKIN1983527, *Bardeen:1983} 
%describes the origin and evolution of our Universe up to the Planck scale.
%The general conclusion of all such experiments is that the inflationary models with a plateau in their field space are now most favored.
However, at the early Universe in the high energy regime ($\sim 10^{16}$ GeV), the gravity as described by Einstein's GR is expected to get modified. One such modification is to consider the higher-order curvature corrections to the Einstein GR, also collectively known as the $f(R)$ gravity \cite{Sotiriou:2008rp, *DeFelice:2010aj}. Most of these modified gravity models entail a coupling between the curvature and scalar field, which mixes the metric and scalar degrees of freedom. This also implies that the effective Planck mass during the early Universe is field-dependent, and hence, time-dependent. Such kind of frames is known as the Jordan frame. It was soon realized that, with a suitable conformal transformation, the extra degree of freedom in those models can be cast into the Einstein theory with a canonical scalar degrees of freedom predicting identical results at the perturbation level, the observable being the scalar spectral tilt, $n_s$ and the scalar-to-tensor ratio, $r$. Thus it is argued that the conformally invariant frames are `equivalent'. However, differences in the dynamics in these two frames has also been pointed out in several articles in the literature \cite{Bezrukov:2009db, *White:2012ya, *Bahamonde:2017kbs, Nandi:2018ooh, *Nandi:2019xlj}.

Reheating is the phase when, after the end of inflation, the inflaton oscillates coherently around the minimum of the potential and decays into other relativistic particles, setting the stage for the radiation dominated era \cite{Albrecht:1982mp, *Abbott:1982hn, *Kofman:1994rk}. The quantities describing the reheating phase depend significantly on the background dynamics, and unlike the perturbations, the background evolution in these two frames are significantly different.  Therefore, there is a possibility that the two frames lead to different reheating dynamics, as implied in Refs. \cite{Nandi:2018ooh, Nandi:2019xlj}.

In this letter, we show that the reheating constraints have indeed the potential to discriminate the two frames. In doing so, we will consider $f(R)$ theories. The inflationary energy scale, as well as the energy density at the end of inflation $\rho_{\rm end}$ in both the frames,  are different. We will see that these differences culminate into the difference in the reheating e-folding number $N_{\rm re}$, and subsequently in the reheating temperature $T_{\rm re}$. Since the tensor modes are yet to be detected, the reheating epoch can be a good measure to distinguish various inflationary models, and in our case, the difference may indeed favor one frame over another.
%We have also seen that the reheating constrains too imply that the Einstein frame description has better agreement with data.

We denote $\mpl \equiv 1/\sqrt{8 \pi G}$ as the reduced Planck mass. Also, all physical quantities in the Einstein frame are denoted with the tilde, \ie $\tilde{X}$, to distinguish the same in the Jordan frame.

\par{\it Governing equations:} 
%\cite{Whitt:1984pd, *Muller:1987hp, *BARROW1988515, *Maeda:1988ab, *Ketov:2012jt, *0022-3689-5-6-005, *Sotiriou:2008rp, *DeFelice:2010aj, *Capozziello:2010zz}.
The action for the $f(R)$ theories is

\begin{eqnarray}\label{Eq:Actionf(R)}
\mathcal{S}_{f(R)} = \frac{M_{\rm pl}^2}{2}\,\int {\rm d}^4 x\,\sqrt{-g}\,f(R),
\end{eqnarray}

\noindent where $R$ is the Ricci scalar. In these theories, along with the transverse tensor degrees of freedom, there is also a scalar degree of freedom $\partial f (R)/\partial R$. Therefore, by changing the field variable from $R$ to $\phi \equiv \partial f (R)/\partial R $, it can be re-written as
 
\begin{eqnarray}\label{Eq:Actionf(R)Jordan}
\mathcal{S}_{f(R)} = \frac{1}{2}\,\int {\rm d}^4 x\,\sqrt{-g}\,\left[ M_{\rm pl}^2 \,\phi\,  R - 2\, V(\phi) \right],
\end{eqnarray}

\noindent where the potential is defined as $V(\phi) = M_{\rm pl}^2/2 \,\left[\phi R\left(\phi\right) - 2f\left(R\left(\phi\right)\right)\right]$. As one can see from the action \eqref{Eq:Actionf(R)Jordan}, the $f(R)$ theory of gravity is identical to the Brans-Dicke theory \cite{Brans-Dicke1961} with the Brans-Dicke parameter $\omega_{\rm BD} = 0$. By using a suitable conformal transformation $\tilde{g}_{\mu \nu} = \phi g_{\mu \nu}$, it can be re-written as

\begin{eqnarray}\label{Eq:ActionEinstein}
\mathcal{S}_{\rm E} = \frac{1}{2}\int d^4{\rm \bf x} \sqrt{-\tilde{g}}\left(M_{\rm pl}^2\,\tilde{R} -   \tilde{g}^{\mu \nu}\tilde{\nabla}_\mu \tilde{\phi}\tilde{\nabla}_\nu \tilde{\phi} -2 \tilde{V}(\tilde{\phi})\right)\nonumber\\
\end{eqnarray}
\noindent in the Einstein frame. The redefined canonical scalar field and the potential in the Einstein frame are defined as
\begin{eqnarray}\label{Eq:Potentialf(R)Einstein}
\phi = \exp\left(\sqrt{\frac{2}{3}}\frac{\tilde{\phi}}{M_{\rm pl}}\right), \tilde{V}(\tilde{\phi}) =  \frac{V(\phi(\tilde{\phi}))}{\phi(\tilde{\phi})^2}.
\end{eqnarray}

\noindent In these frames, the scalar and tensor perturbation observables can easily be obtained by using slow-roll techniques. The scalar spectral index $n_{\rm s}$, and the tensor-to-scalar ratio $r$ are given by the expressions:

\begin{eqnarray}\label{Eq:As}
\label{Eq:ns}
n_{\rm s} &\simeq&
\begin{cases}
1 - 4\,\epsilon_1 + 2\,\epsilon_2 - 2\,\epsilon_3\\
1 - 2\,\tilde{\epsilon}_1 - \tilde{\epsilon}_{2},
\end{cases} \\
\label{Eq:r}
r &\simeq&
\begin{cases}
48\,\epsilon_{3}^2 \\
16\,\tilde{\epsilon_{1}},
\end{cases}
\end{eqnarray}

\noindent where $\{\epsilon_1 \equiv -\dot{H}/H^2, \, \epsilon_{2} = \dot{\phi}/2 H \phi, \, \epsilon_{3} = \ddot{\phi}/H \dot{\phi}\}$ and $\{\tilde{\epsilon}_1 = -\dot{\tilde{H}}/\tilde{H}^2,\, \tilde{\epsilon}_2 = \dot{\tilde{\epsilon}}_1/H \tilde{\epsilon}_1\}$ are the slow-roll parameters in the Jordan and the Einstein frames, respectively.
% and they can be expressed in terms of the potential as

%\begin{eqnarray}
%\epsilon_{V} &\equiv& \frac{M_{\rm p}^2}{2}\left(\frac{V_{\phi}}{V}\right)^{2}, \\
%\eta_{V} & \equiv & M_{\rm p}^2\frac{V_{\phi \phi}}{V}, \\
%\tilde{\epsilon}_{1} & \equiv &\frac{\left(\tphi \tV_{ \tphi}-2 \tV\right)\left(\tphi \tV_{ \tphi}-\tV\right)}{3 \tV^{2}} \\
%\tilde{\epsilon}_{2} &\equiv & \tilde{\epsilon}_{1}-\frac{2 \tphi\left(\tphi \tV_{ \tphi \tphi}-\tV_{\tphi}\right)}{3 \tV} \\
%\tilde{\epsilon}_{3} &\equiv & -\frac{\left(\tphi \tV_{ \tphi}-2 \tV\right)}{3 \tV}.
%\end{eqnarray}

%\noindent Using \eqref{Eq:As}, and \eqref{Eq:StpotJordan} and \eqref{Eq:StpotJordan} in both frames, we can evaluate $m$ as Planck mission suggest $A_{\rm s} \simeq 2.1 \times 10^{-9}$. This implies $m \simeq \tilde{m} \simeq 1.2 \times 10^{-5}$. 
%Also, using \eqref{Eq:ns}, we can evaluate the field $\phi$ and $\tilde{\phi}$ as a function of the spectral index $n_{\rm s}$. In this two frames, they are, $\phi_k \simeq (7 + 4\, \sqrt{4 - 3\, n_{\rm s}} - 3 \,n_{\rm s})/(3 - 3\, n_{\rm s})$ and $\tilde{\phi}_k = \sqrt{3/2}\, \mpl \,{\rm log}\, \phi_{k}$,
%Note that, using above expressions, $r$ in both frames can be expressed as a function of $\ns$. 
It can easily be checked that for the same value of $\ns$, $r$ in both frames remains identical, and thus, inflationary observables are equal, making the frames extremely difficult to distinguish from the CMB observations. However, this is not at all surprising as we know that under conformal transformation, scalar and tensor perturbations remain invariant.

After the end of inflation, the inflaton energy density decreases and the inflaton decays into other relativistic degrees. The expansion of the universe during reheating phase can be parametrized by an effective equation of state parameter $w_{\rm re}$ \cite{Turner:1983he, Dai:2014jja, *Martin:2014nya} such that the energy density during reheating behaves as $\rho \propto a^{-3(1+w_{\rm re})}$. Using this, we first evaluate the reheating e-folding number $N_{\rm re} \equiv \ln(a_{\rm re}/a_{\rm end})$, and consequently the reheating temperature $T_{\rm re} \sim e^{- N_{\rm re} }$ in the respective frames. It can easily be verified that, at the end of inflation, since the conformal factor becomes $\phi \sim 1$, the equations in both frames become nearly similar, and thus the effective equation of state $w_{\rm re}$ in both frames may be considered to be the same. 
%and the range can be $-1/3 \lesssim w_{\rm re} \lesssim 1$.
%Reheating temperature is proportional to the negative exponential of the reheating e-folding, and thus, a slight deviation of reheating e-folding in two frames can lead to different reheating temperature and thus different thermal history of our Universe. In the CMB,  the scales of the perturbed modes are comparable to the horizon size. The corresponding comoving Hubble scale for the mode $k$ leaving the Horizon is $a_k H_k$ and can be related to the present epoch and by using that relation,
The reheating e-folding number can be written as \cite{Liddle:2003as}

\begin{eqnarray}\label{Eq:Nre}
N_{\mathrm{re}} &=& \frac{4}{ 3 w_{\mathrm{re}} - 1}\Bigg[ N_{k} +\ln \left(\frac{k}{a_{0} T_{0}}\right) - \ln \left(H_{k}\right) +\frac{1}{4}\ln(\rho_{\rm end}) \nonumber \\
&&  + \frac{1}{4} \ln \left(\frac{30}{\pi^{2} g_{\rm re}}\right)  + \frac{1}{3} \ln \left(\frac{11 g_{\rm s, re}}{43}\right)\Bigg].
\end{eqnarray}

\noindent $N_k$ is the e-folding number when a particular $k$ mode leaves the Hubble radius to the end of inflation and $\{a_0$, $T_0\}$ are present values of the scale factor and the temperature, respectively. $H_k$ is the Hubble scale when the mode leaves the Horizon; $\rho_{\rm end}$ is the energy density at the end of inflation and $\{g_{\rm re}, g_{\rm s, re}\}$ are the effective number of relativistic species upon thermalization and the effective number of light species for entropy during reheating, respectively.

Here, we work with the pivot scale $k/a_0  = 0.05\, \text{Mpc}^{-1}$ and therefore, the second term in the right hand side is same for both frames. 
%As a consequence, the relation in (\ref{Eq:Nre}) can be simply written as:
%\begin{equation*}
%    N_{\rm re} = \frac{4}{3w_{\rm re} -1}\left[N_k - 61.6 -\ln(H_k) + \frac{1}{4}\ln(\rho_{\rm end})\right]
%\end{equation*}
It can easily be shown that $N_k$ in these two frames are near identical for a fixed value of $n_s$ ($\Delta N_k \sim \mathcal{O}(0.1)$). Therefore, considering ${\{g_{\rm re}, g_{\rm s, re}\}}\sim 100$ in both frames, $N_{\rm re}$ can only differ significantly if there is a difference in $H_k$ and/or $\rho_{\rm end}$ in both frames.

\begin{figure*}[ht]
\centering
\includegraphics[width=0.45\linewidth]{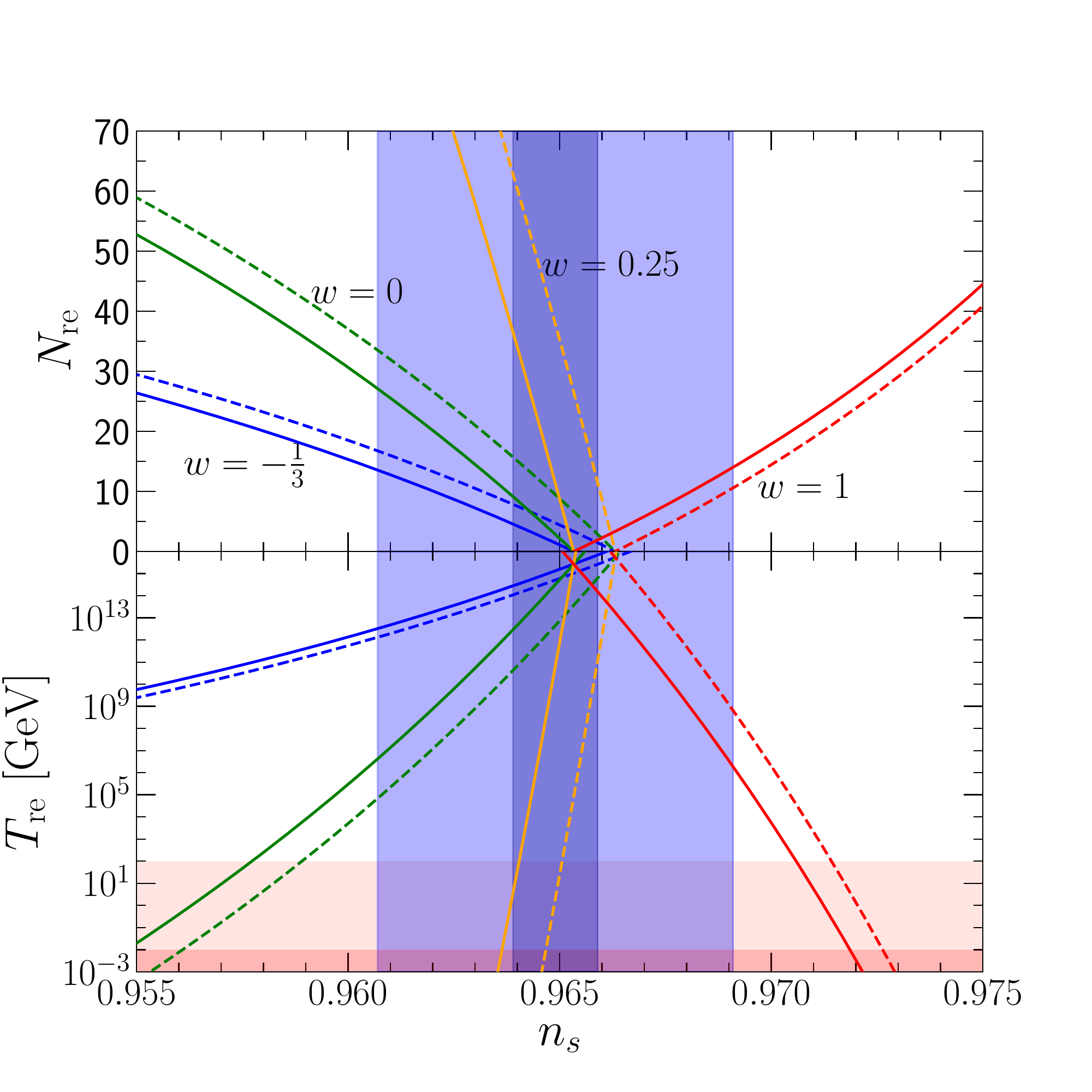}
\includegraphics[width=0.45\linewidth]{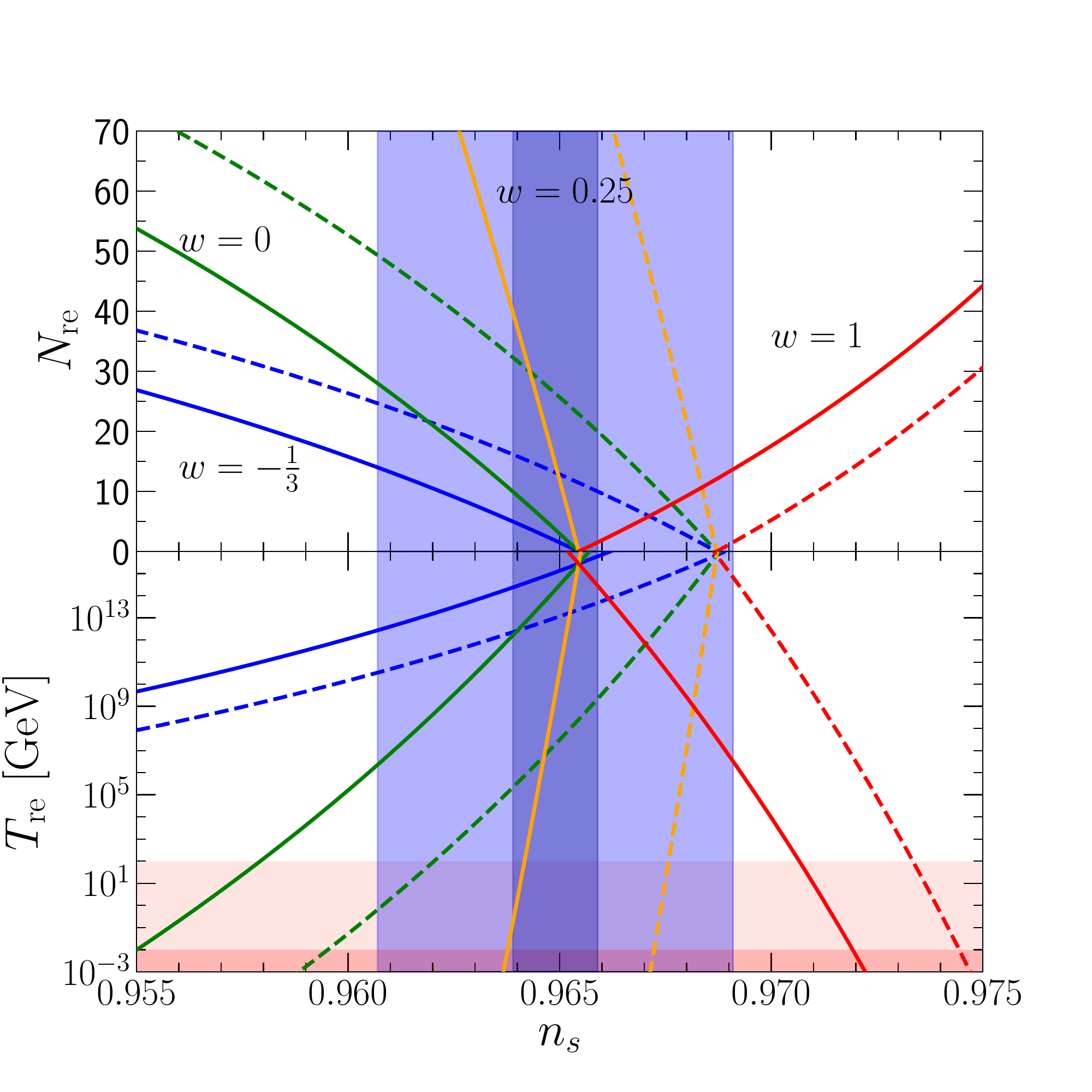}
    \caption{\scriptsize We plot the variation of reheating e-folding number $N_{\rm re}$ and reheating temperature $T_{\rm re}$ with the scalar spectral index $n_s$ for the Starobinsky model (left) and the chaotic $m^2\phi^2$ model (right). The solid lines are for the Einstein frame while the dashed lines are for the Jordan frame. Different colors represent different effective equation of the state $w_{\rm re}$ as indicated in the figure. The blue shaded region is the Planck $1\sigma$ region with $n_s = 0.9649 \pm 0.0042$ \cite{Akrami:2018odb}. The dark blue region indicates the future projected bound on $\ns$ with a sensitivity of $10^{-3}$ with the assumption that the central value of it will remain unchanged \cite{Creminelli:2014fca}. The temperature below the deep red region is excluded due the constraints from BBN \cite{Steigman:2007xt, *Fields:2014uja} while the lighter red region is the electroweak scale taken to be $100$ GeV.}
    \label{Fig:Starobinsky}
\end{figure*}
The Hubble energy scale $H_k$ can be evaluated (using first order slow-roll approximation) and expressed in terms of the amplitude $A_{\rm s}$ of scalar power-spectrum and the tensor-to-scalar ratio $r$ as
\begin{eqnarray}\label{Eq:HdiffObs}
    H_k &\simeq&
    \begin{cases}
     \frac{\pi M_{\rm p}\sqrt{r A_s\phi_{k}}}{\sqrt{2}}\qquad \qquad \text{Jordan frame}\\
    \frac{\pi M_{\rm p}\sqrt{r A_s}}{\sqrt{2}} \qquad \qquad \quad\text{Einstein frame}
    \end{cases}.
\end{eqnarray}
Now, both $\{r, A_{\rm s}\}$ are found to be the same in both frames and therefore, in the Jordan frame, $H_k$ is $\phi_k^{1/2}$ times higher than the same in the Einstein frame. This result is anticipated since the difference in $H_k$ in a conformally related theories with the transformation $\tilde{g}_{\mu \nu} = \phi \,g_{\mu \nu}$ is
\begin{eqnarray}\label{Eq:HdiffConf}
    \tilde{H} = \frac{H}{\sqrt{\phi}} \, (1 + 2 \epsilon_{2}).
\end{eqnarray}
Therefore, during inflation, neglecting the slow-roll parameter, we arrive at $\tilde{H}_k \approx \phi_k{}^{-1/2}\, H_k$. It can also be verified that at this energy scale, all quantities in these frames follow the conformal relation, e.g., $\phi_{k}$ and $\tphi_k$ are related by the conformal relation \eqref{Eq:Potentialf(R)Einstein}.

However, since $\epsilon_1$ is not a conformally invariant quantity, $\epsilon_1 =1$ does not imply $\tilde{\epsilon}_1 = 1$, and therefore quantities at the end of inflation in these frames do not follow the conformal relations, \ie unlike $\phi_{k}$ and $\tphi_k$, $\phi_{\rm end}$ and $\tilde{\phi}_{\rm end}$ does not maintain the conformal relation \eqref{Eq:Potentialf(R)Einstein}. Therefore, there is an inherent difference in $\rho_{\rm end}$ (the effective energy density in the Jordan frame at the end of inflation is $3 H_{\rm end}^2 \mpl^2$) in different frames. The differences in $H_k$ (contributes the most) and $\rho_{\rm end}$ in the respective frames do not compensate and therefore, $N_{\rm re}$ in these two frames differ significantly. As one can readily see from the relation \eqref{Eq:Nre}, for $w_{\rm re} \lesssim 1/3$, in the most plausible scenarios, $N_{\rm re}$ is always higher than $\tilde{N}_{\rm re}$ while the relation reverses for $w_{\rm re} \gtrsim 1/3$.

Assuming the conservation of reheating entropy in the present CMB, the reheating temperature $T_{\rm re}$ can be expressed in terms of $N_{\rm re}$ as \cite{Dai:2014jja, *Martin:2014nya}

\begin{eqnarray}
T_{\rm re} = \left(\frac{43}{11 g_{\rm s, r e}}\right)^{1 / 3}\left(\frac{a_{0} T_{0}}{k}\right) H_k\, e^{-N_k - N_{\rm re}}.
\label{eq:Tre}
\end{eqnarray}
\begin{figure*}[ht!]
    \centering
        \includegraphics[width=0.45\linewidth]{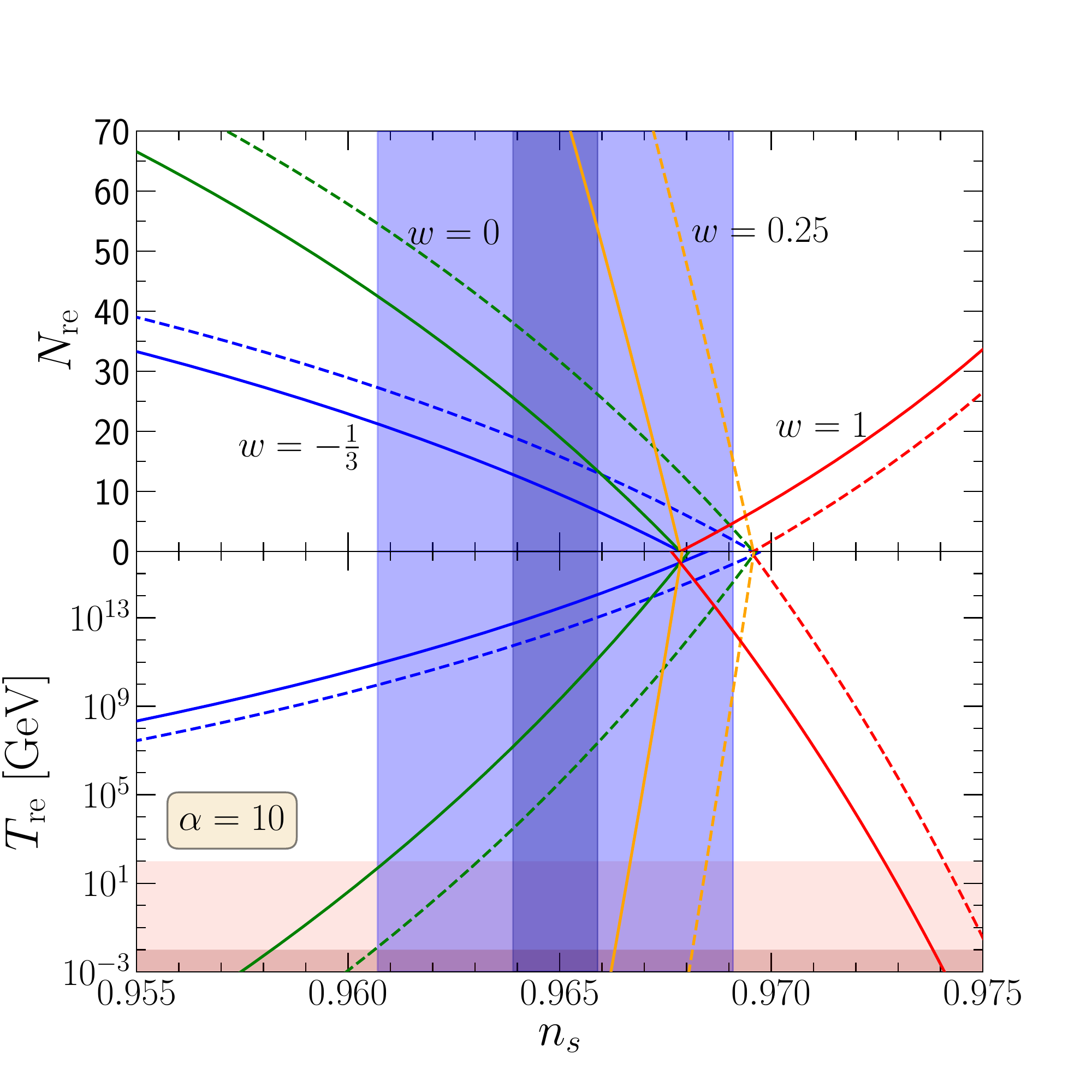}
        \includegraphics[width=0.45\linewidth]{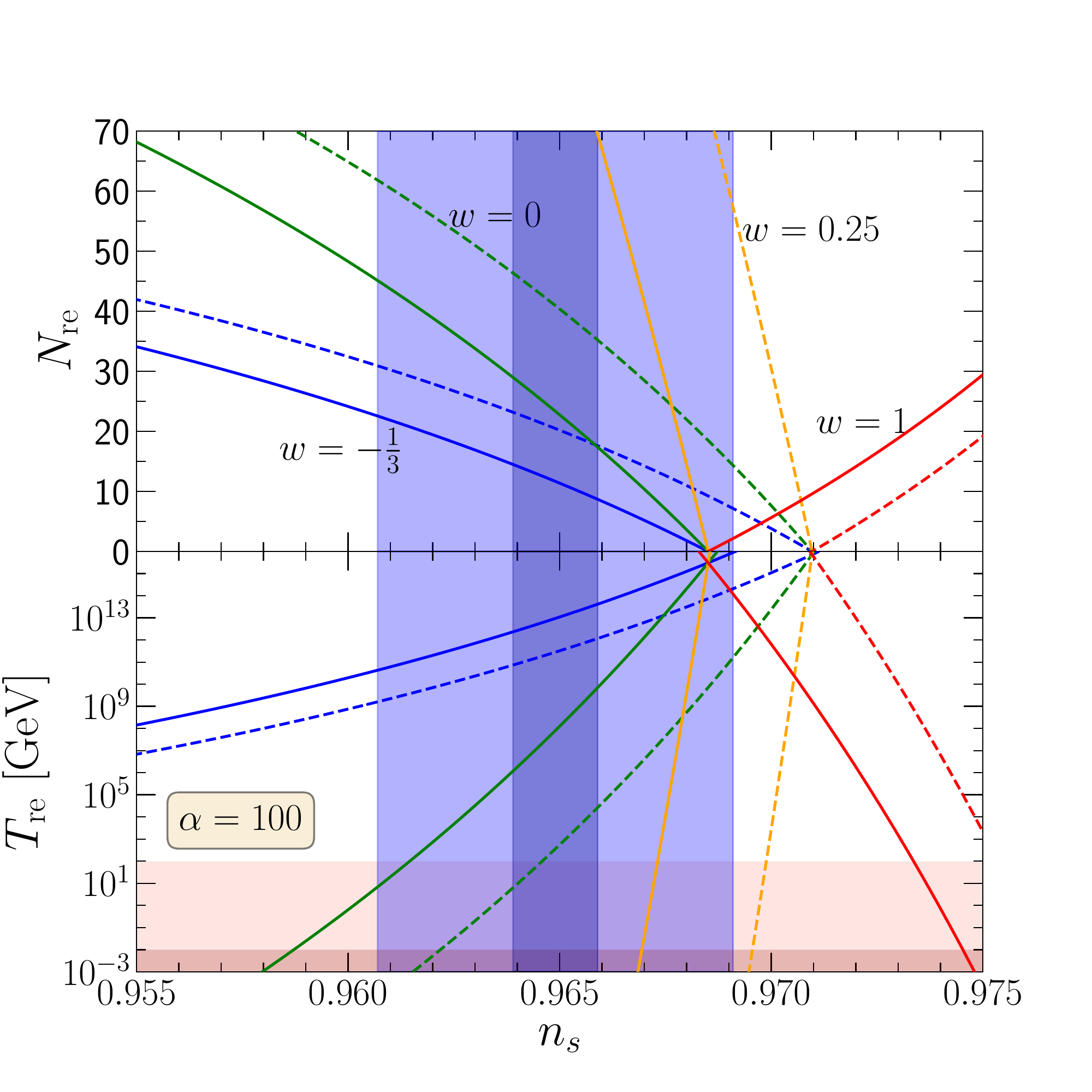}
    \caption{\scriptsize We plot the variation of reheating e-folding number $N_{\rm re}$ and reheating temperature $T_{\rm re}$ with the scalar spectral index $n_s$ for the $\alpha$-attractor model with $\alpha = 10$ (left) and $\alpha= 100$ (right). Note that the Starobinsky model is a special case of $\alpha=1$.}
    \label{Fig:alpha10-100}
\end{figure*}

\noindent As we can see, $T_{\rm re}$ increases with $H_k$ while decreases exponentially with $N_{\rm re}$. Thus, the difference in $H_k$ result in varying the reheating e-folding numbers in the two frames which consequently yields a different reheating temperature. It can also be seen from the relation \eqref{Eq:Nre}, $w_{\rm re}$  plays the role in $N_{\rm re}$ and subsequently, in $T_{\rm re}$ which becomes more prominent as $w_{\rm re}$ approaches $1/3$.

{\it Results:} We now consider specific inflationary models to analyze the difference quantitatively. Arguably, the most favored model in accordance with CMB is the Starobinsky model \cite{STAROBINSKY198099, *STAROBINSKY1982175} with $f(R) = R + 1/(6m^2)\, R^2$ where the Jordan \eqref{Eq:Actionf(R)Jordan} and Einstein frame potential \eqref{Eq:Potentialf(R)Einstein} are $V(\phi) = 3/4\, m^2\, M_{\rm pl}^2\, (1 - \phi)^2$ and $\tilde{V}(\tilde{\phi}) = 3/4\, m^2\,M_{\rm pl}^2\,\left(1 - e^{-\sqrt{2/3} \tilde{\phi}/M_{\rm pl}}\right)^2$. We numerically solve equations \eqref{Eq:Nre} and \eqref{eq:Tre}, and show the difference in $N_{\rm re}$, as well as in $T_{\rm re}$ in both frames for various values of $w_{\rm re}$ ($-1/3 \lesssim w_{\rm re} \lesssim 1$) in the left panel in Fig. \ref{Fig:Starobinsky}. In this case, $\phi_k \sim 10^2$ and therefore, change in $H_k$ is $\sim 10$. Also, by properly evaluating the end of inflation, change in $\rho_{\rm end}$ (higher in the Jordan frame) is $\sim 10$. Thus, the change in $N_{\rm re}$ for $w_{\rm re} \sim 0$ is $ \sim 3 \ln 10$ (higher in the Jordan frame). Therefore, the change in $T_{\rm re}$ for $w_{\rm re} \sim 0$ is $10^{-2}$, \ie the reheating temperature in the Jordan frame is lower than the same in the Einstein frame. We can approximate our numerical result with $w_{\rm re} =0$ by fitting the relation of reheating e-folding and the temperature as a function of $\ns$ as $\{N_{\rm re} \simeq 6425 \,(0.9662 - \ns),\, \log_{10}(T_{\rm re}/ 10^{6} {\rm GeV}) \simeq 2093 \,(\ns - 0.9616) \}$ in the Jordan frame, and $\{\tilde{N}_{\rm re} \simeq 6480\, (0.9652 - x), \, \log_{10}(\tilde{T}_{\rm re}/ 10^{6} {\rm GeV}) \simeq 2111\, (\ns - 0.9608)$ in the Einstein frame. The difference in reheating temperature is apparent from these relations.
%\begin{figure}
%    \includegraphics[scale=0.4]{chaotic}
%    \caption{\scriptsize We have plotted the variation of reheating e-folding number($N_{\rm re}$) and reheating temperature($T_{\rm re}$) with the scalar spectral index $n_s$  for chaotic $m^2\phi^2$ model. The solid lines are for Einstein frame while the dashed lines are for Jordan frame. As one can see, in this case, the difference is more prominent than the Starobinsky model.}
%    \label{Fig:Chaotic}
%\end{figure}

In the case of large field models, the difference can be more prominent as it allows $\phi_{k}$ to be very high. In the right panel of Fig. \ref{Fig:Starobinsky}, we plot the same for the chaotic inflation model \cite{Linde:1983gd}: a large field model. In the Einstein frame, the potential is $\tV(\tphi) = 1/2 m^2 \,\tphi{}^2$ and by using \eqref{Eq:Potentialf(R)Einstein}, the potential in the Jordan frame becomes $V(\phi) = 3/4 \,m^2 \mpl^2\, (\phi \ln \phi)^2.$  In this case, $\phi_{k} \sim 10^5$ is significantly higher than the same in the Starobinsky model. This reflects in difference in reheating temperatures $\sim 10^8$.
In Fig. \ref{Fig:alpha10-100}, we also plot the differences in two frames for two $\alpha$-attractor models \cite{Galante:2014ifa, *Kallosh:2013tua, *Kallosh:2013yoa} (note that, $\alpha = 1$ is the Starobinsky model) with $\alpha = 10$ and $100$. Since $\phi_{k}$ increases with $\alpha$, therefore, the difference becomes more prominent with increased value of $\alpha$.

In Fig. \ref{Fig:Starobinsky}, as one can see, the additional reheating constraint is the upper bound (lower bound for $w_{\rm re} \gtrsim 1/3$) on $\ns$ since $N_{\rm re}$ must be positive. In the Einstein frame, $\ns \lesssim 0.9652$, whereas, in the Jordan frame, the bound is $\ns \lesssim 0.9662$, \ie $\Delta \ns \sim 10^{-3}$. In the case of chaotic inflation (large field models as well as with increasing $\alpha$ in the $\alpha$-attractor model), $\Delta \ns$ increases (see Figs. \ref{Fig:Starobinsky} and \ref{Fig:alpha10-100}). From these figures, we can easily infer that in the Jordan frame, the upper bound is always higher than the same in the Einstein frame. 

{\it Discussion and conclusions:} In this letter, we considered the most conservative allowed range of the effective equation of state as $-1/3 \lesssim w_{\rm re} \lesssim 1$. However, at the end of inflation, the equation of state is $-1/3$ whereas $w_{\rm re} \gtrsim  1/3$ is difficult to conceive since it requires a potential dominated by higher-dimension operators (higher than $\phi^6$) near its minimum. Therefore, the plausible range of $w_{\rm re}$ is $-1/3 \lesssim w_{\rm re} \lesssim 1/3$. In addition to that, preheating study performed with the help of lattice simulations shows that the equation of state $w_{\rm re}$ during this phase is not a constant but varies smoothly in the range of  $0 \lesssim w_{\rm re} \lesssim 0.25$ while the duration is nearly instantaneous, \ie $N_{\rm re} \sim 0$ \cite{Podolsky:2005bw}. Therefore, considering instantaneous reheating as a benchmark and with the help of future experiments such as EUCLID \cite{Amendola:2012ys} and PRISM \cite{Andre:2013afa}, cosmic 21-cm surveys \cite{Mao:2008ug} and CORE experiments\cite{Finelli:2016cyd} with improved precision of $10^{-3}$ in $\ns$, any model in these frames can be tightly constrained. For instance, the Starobinsky model in the Jordan frame is in tension as it is just outside the future constrained region (see the left panel in Fig. \ref{Fig:Starobinsky}). Also, there is a tighter constraint on the maximum value of $\alpha$ in the $\alpha$-attractor model (see Fig. \ref{Fig:alpha10-100}). Besides, the primordial gravitational waves provide a compelling window to look at physics beyond the BBN and thereby may constrain the reheating phase \cite{Boyle:2007zx,*Koh:2018qcy}.  %The experiments such as LIGO/Virgo\cite{Abbott:2017xzu} or proposed future detectors such as LISA\cite{Audley:2017drz}, BBO/DECIGO\cite{Yagi:2011wg}, the Einstein Telescope (ET)\cite{Punturo:2010zz}, and Cosmic Explorer (CE)\cite{Evans:2016mbw} will eventually be able to probe gravitational waves emitted during the period of inflation.

The conformally connected frames, which was previously thought to be `equivalent', can now be distinguished if we take the reheating constraints into account. We showed that, because of the conformal factor, the inflationary energy scale differs in different frames. In addition to that, at the end of inflation, the energy density in both frames differs. All these culminate into different values of reheating e-folding number $N_{\rm re}$. This difference, in turn, causes a significant change in reheating temperature $T_{\rm re}$ and subsequently, different reheating constraints. In case of small field models, the difference in $T_{\rm re}$ as well as the difference in the upper bound on $\ns$ in both frames are small compared to the same in the large field models. These results not only breaks the idea of similarity in physics in conformally connected frames, as we thought them to be, but also provide viable observational bounds that can be verified with future observations.

%The application of this result is that the non-minimal and conformally coupled minimal theory can now be treated as different theories.

Using similar arguments, we expect the difference in reheating dynamics in degenerate theories (where inflationary observables are identical). One such example is the Starobinsky model and the Higgs inflation. This work is under investigation.
%that, we can also establish the difference in Higgs inflation and Starobinsky inflation as they are conformally related. Also, we are currently focusing on qualitative reheating analysis. These works are under progress.

\par{\it Acknowledgements:} We thank L. Sriramkumar for useful discussions and his valuable comments on this work. We also thank the Indian Institute of Technology Madras, Chennai, India for the support through Institute Postdoctoral fellowship. We wish to acknowledge partial support from the Science and Engineering Research Board, Department of Science and Technology, Government of India, through the Core Research Grant CRG/2018/002200.


%%%%%%%%%%%%%%%%%%%%%%%%%%%%%%%%%%%%%%%%%%%%%%%%%%%%%%%%%%%%%%%%%%%%%%%%%%%%%%%


\begin{thebibliography}{34}%
	\makeatletter
	\providecommand \@ifxundefined [1]{%
		\@ifx{#1\undefined}
	}%
	\providecommand \@ifnum [1]{%
		\ifnum #1\expandafter \@firstoftwo
		\else \expandafter \@secondoftwo
		\fi
	}%
	\providecommand \@ifx [1]{%
		\ifx #1\expandafter \@firstoftwo
		\else \expandafter \@secondoftwo
		\fi
	}%
	\providecommand \natexlab [1]{#1}%
	\providecommand \enquote  [1]{``#1''}%
	\providecommand \bibnamefont  [1]{#1}%
	\providecommand \bibfnamefont [1]{#1}%
	\providecommand \citenamefont [1]{#1}%
	\providecommand \href@noop [0]{\@secondoftwo}%
	\providecommand \href [0]{\begingroup \@sanitize@url \@href}%
	\providecommand \@href[1]{\@@startlink{#1}\@@href}%
	\providecommand \@@href[1]{\endgroup#1\@@endlink}%
	\providecommand \@sanitize@url [0]{\catcode `\\12\catcode `\$12\catcode
		`\&12\catcode `\#12\catcode `\^12\catcode `\_12\catcode `\%12\relax}%
	\providecommand \@@startlink[1]{}%
	\providecommand \@@endlink[0]{}%
	\providecommand \url  [0]{\begingroup\@sanitize@url \@url }%
	\providecommand \@url [1]{\endgroup\@href {#1}{\urlprefix }}%
	\providecommand \urlprefix  [0]{URL }%
	\providecommand \Eprint [0]{\href }%
	\providecommand \doibase [0]{http://dx.doi.org/}%
	\providecommand \selectlanguage [0]{\@gobble}%
	\providecommand \bibinfo  [0]{\@secondoftwo}%
	\providecommand \bibfield  [0]{\@secondoftwo}%
	\providecommand \translation [1]{[#1]}%
	\providecommand \BibitemOpen [0]{}%
	\providecommand \bibitemStop [0]{}%
	\providecommand \bibitemNoStop [0]{.\EOS\space}%
	\providecommand \EOS [0]{\spacefactor3000\relax}%
	\providecommand \BibitemShut  [1]{\csname bibitem#1\endcsname}%
	\let\auto@bib@innerbib\@empty
	%</preamble>
	\bibitem [{\citenamefont {Aghanim}\ \emph {et~al.}(2018)\citenamefont {Aghanim}
		\emph {et~al.}}]{Aghanim:2018eyx}%
	\BibitemOpen
	\bibfield  {author} {\bibinfo {author} {\bibfnamefont {N.}~\bibnamefont
			{Aghanim}} \emph {et~al.} (\bibinfo {collaboration} {Planck}),\ }\href@noop
	{} {\  (\bibinfo {year} {2018})},\ \Eprint {http://arxiv.org/abs/1807.06209}
	{arXiv:1807.06209 [astro-ph.CO]} \BibitemShut {NoStop}%
	%%CITATION = ARXIV:1807.06209;%%
	\bibitem [{\citenamefont {Abbott}\ \emph {et~al.}(2019)\citenamefont {Abbott}
		\emph {et~al.}}]{Abbott:2018lct}%
	\BibitemOpen
	\bibfield  {author} {\bibinfo {author} {\bibfnamefont {B.~P.}\ \bibnamefont
			{Abbott}} \emph {et~al.} (\bibinfo {collaboration} {LIGO Scientific,
			Virgo}),\ }\href {\doibase 10.1103/PhysRevLett.123.011102} {\bibfield
		{journal} {\bibinfo  {journal} {Phys. Rev. Lett.}\ }\textbf {\bibinfo
			{volume} {123}},\ \bibinfo {pages} {011102} (\bibinfo {year} {2019})},\
	\Eprint {http://arxiv.org/abs/1811.00364} {arXiv:1811.00364 [gr-qc]}
	\BibitemShut {NoStop}%
	%%CITATION = ARXIV:1811.00364;%%
	\bibitem [{\citenamefont {Akrami}\ \emph {et~al.}(2018)\citenamefont {Akrami}
		\emph {et~al.}}]{Akrami:2018odb}%
	\BibitemOpen
	\bibfield  {author} {\bibinfo {author} {\bibfnamefont {Y.}~\bibnamefont
			{Akrami}} \emph {et~al.} (\bibinfo {collaboration} {Planck}),\ }\href@noop {}
	{\  (\bibinfo {year} {2018})},\ \Eprint {http://arxiv.org/abs/1807.06211}
	{arXiv:1807.06211 [astro-ph.CO]} \BibitemShut {NoStop}%
	%%CITATION = ARXIV:1807.06211;%%
	\bibitem [{\citenamefont {Sotiriou}\ and\ \citenamefont
		{Faraoni}(2010)}]{Sotiriou:2008rp}%
	\BibitemOpen
	\bibfield  {author} {\bibinfo {author} {\bibfnamefont {T.~P.}\ \bibnamefont
			{Sotiriou}}\ and\ \bibinfo {author} {\bibfnamefont {V.}~\bibnamefont
			{Faraoni}},\ }\href {\doibase 10.1103/RevModPhys.82.451} {\bibfield
		{journal} {\bibinfo  {journal} {Rev. Mod. Phys.}\ }\textbf {\bibinfo {volume}
			{82}},\ \bibinfo {pages} {451} (\bibinfo {year} {2010})},\ \Eprint
	{http://arxiv.org/abs/0805.1726} {arXiv:0805.1726 [gr-qc]} \BibitemShut
	{NoStop}%
	%%CITATION = ARXIV:0805.1726;%%
	\bibitem [{\citenamefont {De~Felice}\ and\ \citenamefont
		{Tsujikawa}(2010)}]{DeFelice:2010aj}%
	\BibitemOpen
	\bibfield  {author} {\bibinfo {author} {\bibfnamefont {A.}~\bibnamefont
			{De~Felice}}\ and\ \bibinfo {author} {\bibfnamefont {S.}~\bibnamefont
			{Tsujikawa}},\ }\href {\doibase 10.12942/lrr-2010-3} {\bibfield  {journal}
		{\bibinfo  {journal} {Living Rev. Rel.}\ }\textbf {\bibinfo {volume} {13}},\
		\bibinfo {pages} {3} (\bibinfo {year} {2010})},\ \Eprint
	{http://arxiv.org/abs/1002.4928} {arXiv:1002.4928 [gr-qc]} \BibitemShut
	{NoStop}%
	%%CITATION = ARXIV:1002.4928;%%
	\bibitem [{\citenamefont {Bezrukov}\ and\ \citenamefont
		{Shaposhnikov}(2009)}]{Bezrukov:2009db}%
	\BibitemOpen
	\bibfield  {author} {\bibinfo {author} {\bibfnamefont {F.}~\bibnamefont
			{Bezrukov}}\ and\ \bibinfo {author} {\bibfnamefont {M.}~\bibnamefont
			{Shaposhnikov}},\ }\href {\doibase 10.1088/1126-6708/2009/07/089} {\bibfield
		{journal} {\bibinfo  {journal} {JHEP}\ }\textbf {\bibinfo {volume} {07}},\
		\bibinfo {pages} {089} (\bibinfo {year} {2009})},\ \Eprint
	{http://arxiv.org/abs/0904.1537} {arXiv:0904.1537 [hep-ph]} \BibitemShut
	{NoStop}%
	%%CITATION = ARXIV:0904.1537;%%
	\bibitem [{\citenamefont {White}\ \emph {et~al.}(2012)\citenamefont {White},
		\citenamefont {Minamitsuji},\ and\ \citenamefont {Sasaki}}]{White:2012ya}%
	\BibitemOpen
	\bibfield  {author} {\bibinfo {author} {\bibfnamefont {J.}~\bibnamefont
			{White}}, \bibinfo {author} {\bibfnamefont {M.}~\bibnamefont {Minamitsuji}},
		\ and\ \bibinfo {author} {\bibfnamefont {M.}~\bibnamefont {Sasaki}},\ }\href
	{\doibase 10.1088/1475-7516/2012/07/039} {\bibfield  {journal} {\bibinfo
			{journal} {JCAP}\ }\textbf {\bibinfo {volume} {1207}},\ \bibinfo {pages}
		{039} (\bibinfo {year} {2012})},\ \Eprint {http://arxiv.org/abs/1205.0656}
	{arXiv:1205.0656 [astro-ph.CO]} \BibitemShut {NoStop}%
	%%CITATION = ARXIV:1205.0656;%%
	\bibitem [{\citenamefont {Bahamonde}\ \emph {et~al.}(2017)\citenamefont
		{Bahamonde}, \citenamefont {Odintsov}, \citenamefont {Oikonomou},\ and\
		\citenamefont {Tretyakov}}]{Bahamonde:2017kbs}%
	\BibitemOpen
	\bibfield  {author} {\bibinfo {author} {\bibfnamefont {S.}~\bibnamefont
			{Bahamonde}}, \bibinfo {author} {\bibfnamefont {S.~D.}\ \bibnamefont
			{Odintsov}}, \bibinfo {author} {\bibfnamefont {V.~K.}\ \bibnamefont
			{Oikonomou}}, \ and\ \bibinfo {author} {\bibfnamefont {P.~V.}\ \bibnamefont
			{Tretyakov}},\ }\href {\doibase 10.1016/j.physletb.2017.01.012} {\bibfield
		{journal} {\bibinfo  {journal} {Phys. Lett.}\ }\textbf {\bibinfo {volume}
			{B766}},\ \bibinfo {pages} {225} (\bibinfo {year} {2017})},\ \Eprint
	{http://arxiv.org/abs/1701.02381} {arXiv:1701.02381 [gr-qc]} \BibitemShut
	{NoStop}%
	%%CITATION = ARXIV:1701.02381;%%
	\bibitem [{\citenamefont {Nandi}(2018)}]{Nandi:2018ooh}%
	\BibitemOpen
	\bibfield  {author} {\bibinfo {author} {\bibfnamefont {D.}~\bibnamefont
			{Nandi}},\ }\href {\doibase 10.1088/1475-7516/2019/05/040} {\bibfield
		{journal} {\bibinfo  {journal} {JCAP}\ }\textbf {\bibinfo {volume} {1905}},\
		\bibinfo {pages} {040} (\bibinfo {year} {2018})},\ \Eprint
	{http://arxiv.org/abs/1811.09625} {arXiv:1811.09625 [gr-qc]} \BibitemShut
	{NoStop}%
	%%CITATION = ARXIV:1811.09625;%%
	\bibitem [{\citenamefont {Nandi}(2019)}]{Nandi:2019xlj}%
	\BibitemOpen
	\bibfield  {author} {\bibinfo {author} {\bibfnamefont {D.}~\bibnamefont
			{Nandi}},\ }\href {\doibase 10.1103/PhysRevD.99.103532} {\bibfield  {journal}
		{\bibinfo  {journal} {Phys. Rev.}\ }\textbf {\bibinfo {volume} {D99}},\
		\bibinfo {pages} {103532} (\bibinfo {year} {2019})},\ \Eprint
	{http://arxiv.org/abs/1904.00153} {arXiv:1904.00153 [gr-qc]} \BibitemShut
	{NoStop}%
	%%CITATION = ARXIV:1904.00153;%%
	\bibitem [{\citenamefont {Albrecht}\ \emph {et~al.}(1982)\citenamefont
		{Albrecht}, \citenamefont {Steinhardt}, \citenamefont {Turner},\ and\
		\citenamefont {Wilczek}}]{Albrecht:1982mp}%
	\BibitemOpen
	\bibfield  {author} {\bibinfo {author} {\bibfnamefont {A.}~\bibnamefont
			{Albrecht}}, \bibinfo {author} {\bibfnamefont {P.~J.}\ \bibnamefont
			{Steinhardt}}, \bibinfo {author} {\bibfnamefont {M.~S.}\ \bibnamefont
			{Turner}}, \ and\ \bibinfo {author} {\bibfnamefont {F.}~\bibnamefont
			{Wilczek}},\ }\href {\doibase 10.1103/PhysRevLett.48.1437} {\bibfield
		{journal} {\bibinfo  {journal} {Phys. Rev. Lett.}\ }\textbf {\bibinfo
			{volume} {48}},\ \bibinfo {pages} {1437} (\bibinfo {year}
		{1982})}\BibitemShut {NoStop}%
	%%CITATION = PRLTA,48,1437;%%
	\bibitem [{\citenamefont {Abbott}\ \emph {et~al.}(1982)\citenamefont {Abbott},
		\citenamefont {Farhi},\ and\ \citenamefont {Wise}}]{Abbott:1982hn}%
	\BibitemOpen
	\bibfield  {author} {\bibinfo {author} {\bibfnamefont {L.~F.}\ \bibnamefont
			{Abbott}}, \bibinfo {author} {\bibfnamefont {E.}~\bibnamefont {Farhi}}, \
		and\ \bibinfo {author} {\bibfnamefont {M.~B.}\ \bibnamefont {Wise}},\ }\href
	{\doibase 10.1016/0370-2693(82)90867-X} {\bibfield  {journal} {\bibinfo
			{journal} {Phys. Lett.}\ }\textbf {\bibinfo {volume} {117B}},\ \bibinfo
		{pages} {29} (\bibinfo {year} {1982})}\BibitemShut {NoStop}%
	%%CITATION = PHLTA,117B,29;%%
	\bibitem [{\citenamefont {Kofman}\ \emph {et~al.}(1994)\citenamefont {Kofman},
		\citenamefont {Linde},\ and\ \citenamefont {Starobinsky}}]{Kofman:1994rk}%
	\BibitemOpen
	\bibfield  {author} {\bibinfo {author} {\bibfnamefont {L.}~\bibnamefont
			{Kofman}}, \bibinfo {author} {\bibfnamefont {A.~D.}\ \bibnamefont {Linde}}, \
		and\ \bibinfo {author} {\bibfnamefont {A.~A.}\ \bibnamefont {Starobinsky}},\
	}\href {\doibase 10.1103/PhysRevLett.73.3195} {\bibfield  {journal} {\bibinfo
			{journal} {Phys. Rev. Lett.}\ }\textbf {\bibinfo {volume} {73}},\ \bibinfo
		{pages} {3195} (\bibinfo {year} {1994})},\ \Eprint
	{http://arxiv.org/abs/hep-th/9405187} {arXiv:hep-th/9405187 [hep-th]}
	\BibitemShut {NoStop}%
	%%CITATION = HEP-TH/9405187;%%
	\bibitem [{\citenamefont {Brans}\ and\ \citenamefont
		{Dicke}(1961)}]{Brans-Dicke1961}%
	\BibitemOpen
	\bibfield  {author} {\bibinfo {author} {\bibfnamefont {C.}~\bibnamefont
			{Brans}}\ and\ \bibinfo {author} {\bibfnamefont {R.~H.}\ \bibnamefont
			{Dicke}},\ }\href {\doibase 10.1103/PhysRev.124.925} {\bibfield  {journal}
		{\bibinfo  {journal} {Phys. Rev.}\ }\textbf {\bibinfo {volume} {124}},\
		\bibinfo {pages} {925} (\bibinfo {year} {1961})}\BibitemShut {NoStop}%
	\bibitem [{\citenamefont {Turner}(1983)}]{Turner:1983he}%
	\BibitemOpen
	\bibfield  {author} {\bibinfo {author} {\bibfnamefont {M.~S.}\ \bibnamefont
			{Turner}},\ }\href {\doibase 10.1103/PhysRevD.28.1243} {\bibfield  {journal}
		{\bibinfo  {journal} {Phys. Rev.}\ }\textbf {\bibinfo {volume} {D28}},\
		\bibinfo {pages} {1243} (\bibinfo {year} {1983})}\BibitemShut {NoStop}%
	%%CITATION = PHRVA,D28,1243;%%
	\bibitem [{\citenamefont {Dai}\ \emph {et~al.}(2014)\citenamefont {Dai},
		\citenamefont {Kamionkowski},\ and\ \citenamefont {Wang}}]{Dai:2014jja}%
	\BibitemOpen
	\bibfield  {author} {\bibinfo {author} {\bibfnamefont {L.}~\bibnamefont
			{Dai}}, \bibinfo {author} {\bibfnamefont {M.}~\bibnamefont {Kamionkowski}}, \
		and\ \bibinfo {author} {\bibfnamefont {J.}~\bibnamefont {Wang}},\ }\href
	{\doibase 10.1103/PhysRevLett.113.041302} {\bibfield  {journal} {\bibinfo
			{journal} {Phys. Rev. Lett.}\ }\textbf {\bibinfo {volume} {113}},\ \bibinfo
		{pages} {041302} (\bibinfo {year} {2014})},\ \Eprint
	{http://arxiv.org/abs/1404.6704} {arXiv:1404.6704 [astro-ph.CO]} \BibitemShut
	{NoStop}%
	%%CITATION = ARXIV:1404.6704;%%
	\bibitem [{\citenamefont {Martin}\ \emph {et~al.}(2015)\citenamefont {Martin},
		\citenamefont {Ringeval},\ and\ \citenamefont {Vennin}}]{Martin:2014nya}%
	\BibitemOpen
	\bibfield  {author} {\bibinfo {author} {\bibfnamefont {J.}~\bibnamefont
			{Martin}}, \bibinfo {author} {\bibfnamefont {C.}~\bibnamefont {Ringeval}}, \
		and\ \bibinfo {author} {\bibfnamefont {V.}~\bibnamefont {Vennin}},\ }\href
	{\doibase 10.1103/PhysRevLett.114.081303} {\bibfield  {journal} {\bibinfo
			{journal} {Phys. Rev. Lett.}\ }\textbf {\bibinfo {volume} {114}},\ \bibinfo
		{pages} {081303} (\bibinfo {year} {2015})},\ \Eprint
	{http://arxiv.org/abs/1410.7958} {arXiv:1410.7958 [astro-ph.CO]} \BibitemShut
	{NoStop}%
	%%CITATION = ARXIV:1410.7958;%%
	\bibitem [{\citenamefont {Liddle}\ and\ \citenamefont
		{Leach}(2003)}]{Liddle:2003as}%
	\BibitemOpen
	\bibfield  {author} {\bibinfo {author} {\bibfnamefont {A.~R.}\ \bibnamefont
			{Liddle}}\ and\ \bibinfo {author} {\bibfnamefont {S.~M.}\ \bibnamefont
			{Leach}},\ }\href {\doibase 10.1103/PhysRevD.68.103503} {\bibfield  {journal}
		{\bibinfo  {journal} {Phys. Rev.}\ }\textbf {\bibinfo {volume} {D68}},\
		\bibinfo {pages} {103503} (\bibinfo {year} {2003})},\ \Eprint
	{http://arxiv.org/abs/astro-ph/0305263} {arXiv:astro-ph/0305263 [astro-ph]}
	\BibitemShut {NoStop}%
	%%CITATION = ASTRO-PH/0305263;%%
	\bibitem [{\citenamefont {Creminelli}\ \emph {et~al.}(2014)\citenamefont
		{Creminelli}, \citenamefont {López~Nacir}, \citenamefont {Simonović},
		\citenamefont {Trevisan},\ and\ \citenamefont
		{Zaldarriaga}}]{Creminelli:2014fca}%
	\BibitemOpen
	\bibfield  {author} {\bibinfo {author} {\bibfnamefont {P.}~\bibnamefont
			{Creminelli}}, \bibinfo {author} {\bibfnamefont {D.}~\bibnamefont
			{López~Nacir}}, \bibinfo {author} {\bibfnamefont {M.}~\bibnamefont
			{Simonović}}, \bibinfo {author} {\bibfnamefont {G.}~\bibnamefont
			{Trevisan}}, \ and\ \bibinfo {author} {\bibfnamefont {M.}~\bibnamefont
			{Zaldarriaga}},\ }\href {\doibase 10.1103/PhysRevD.90.083513} {\bibfield
		{journal} {\bibinfo  {journal} {Phys. Rev.}\ }\textbf {\bibinfo {volume}
			{D90}},\ \bibinfo {pages} {083513} (\bibinfo {year} {2014})},\ \Eprint
	{http://arxiv.org/abs/1405.6264} {arXiv:1405.6264 [astro-ph.CO]} \BibitemShut
	{NoStop}%
	%%CITATION = ARXIV:1405.6264;%%
	\bibitem [{\citenamefont {Steigman}(2007)}]{Steigman:2007xt}%
	\BibitemOpen
	\bibfield  {author} {\bibinfo {author} {\bibfnamefont {G.}~\bibnamefont
			{Steigman}},\ }\href {\doibase 10.1146/annurev.nucl.56.080805.140437}
	{\bibfield  {journal} {\bibinfo  {journal} {Ann. Rev. Nucl. Part. Sci.}\
		}\textbf {\bibinfo {volume} {57}},\ \bibinfo {pages} {463} (\bibinfo {year}
		{2007})},\ \Eprint {http://arxiv.org/abs/0712.1100} {arXiv:0712.1100
		[astro-ph]} \BibitemShut {NoStop}%
	%%CITATION = ARXIV:0712.1100;%%
	\bibitem [{\citenamefont {Fields}\ \emph {et~al.}(2014)\citenamefont {Fields},
		\citenamefont {Molaro},\ and\ \citenamefont {Sarkar}}]{Fields:2014uja}%
	\BibitemOpen
	\bibfield  {author} {\bibinfo {author} {\bibfnamefont {B.~D.}\ \bibnamefont
			{Fields}}, \bibinfo {author} {\bibfnamefont {P.}~\bibnamefont {Molaro}}, \
		and\ \bibinfo {author} {\bibfnamefont {S.}~\bibnamefont {Sarkar}},\
	}\href@noop {} {\bibfield  {journal} {\bibinfo  {journal} {Chin. Phys.}\
		}\textbf {\bibinfo {volume} {C38}},\ \bibinfo {pages} {339} (\bibinfo {year}
		{2014})},\ \Eprint {http://arxiv.org/abs/1412.1408} {arXiv:1412.1408
		[astro-ph.CO]} \BibitemShut {NoStop}%
	%%CITATION = ARXIV:1412.1408;%%
	\bibitem [{\citenamefont {Starobinsky}(1980)}]{STAROBINSKY198099}%
	\BibitemOpen
	\bibfield  {author} {\bibinfo {author} {\bibfnamefont {A.}~\bibnamefont
			{Starobinsky}},\ }\href {\doibase
		https://doi.org/10.1016/0370-2693(80)90670-X} {\bibfield  {journal} {\bibinfo
			{journal} {Physics Letters B}\ }\textbf {\bibinfo {volume} {91}},\ \bibinfo
		{pages} {99 } (\bibinfo {year} {1980})}\BibitemShut {NoStop}%
	\bibitem [{\citenamefont {Starobinsky}(1982)}]{STAROBINSKY1982175}%
	\BibitemOpen
	\bibfield  {author} {\bibinfo {author} {\bibfnamefont {A.}~\bibnamefont
			{Starobinsky}},\ }\href {\doibase
		http://dx.doi.org/10.1016/0370-2693(82)90541-X} {\bibfield  {journal}
		{\bibinfo  {journal} {Physics Letters B}\ }\textbf {\bibinfo {volume}
			{117}},\ \bibinfo {pages} {175 } (\bibinfo {year} {1982})}\BibitemShut
	{NoStop}%
	\bibitem [{\citenamefont {Linde}(1983)}]{Linde:1983gd}%
	\BibitemOpen
	\bibfield  {author} {\bibinfo {author} {\bibfnamefont {A.~D.}\ \bibnamefont
			{Linde}},\ }\href {\doibase 10.1016/0370-2693(83)90837-7} {\bibfield
		{journal} {\bibinfo  {journal} {Phys. Lett.}\ }\textbf {\bibinfo {volume}
			{129B}},\ \bibinfo {pages} {177} (\bibinfo {year} {1983})}\BibitemShut
	{NoStop}%
	%%CITATION = PHLTA,129B,177;%%
	\bibitem [{\citenamefont {Galante}\ \emph {et~al.}(2015)\citenamefont
		{Galante}, \citenamefont {Kallosh}, \citenamefont {Linde},\ and\
		\citenamefont {Roest}}]{Galante:2014ifa}%
	\BibitemOpen
	\bibfield  {author} {\bibinfo {author} {\bibfnamefont {M.}~\bibnamefont
			{Galante}}, \bibinfo {author} {\bibfnamefont {R.}~\bibnamefont {Kallosh}},
		\bibinfo {author} {\bibfnamefont {A.}~\bibnamefont {Linde}}, \ and\ \bibinfo
		{author} {\bibfnamefont {D.}~\bibnamefont {Roest}},\ }\href {\doibase
		10.1103/PhysRevLett.114.141302} {\bibfield  {journal} {\bibinfo  {journal}
			{Phys. Rev. Lett.}\ }\textbf {\bibinfo {volume} {114}},\ \bibinfo {pages}
		{141302} (\bibinfo {year} {2015})},\ \Eprint {http://arxiv.org/abs/1412.3797}
	{arXiv:1412.3797 [hep-th]} \BibitemShut {NoStop}%
	%%CITATION = ARXIV:1412.3797;%%
	\bibitem [{\citenamefont {Kallosh}\ \emph {et~al.}(2014)\citenamefont
		{Kallosh}, \citenamefont {Linde},\ and\ \citenamefont
		{Roest}}]{Kallosh:2013tua}%
	\BibitemOpen
	\bibfield  {author} {\bibinfo {author} {\bibfnamefont {R.}~\bibnamefont
			{Kallosh}}, \bibinfo {author} {\bibfnamefont {A.}~\bibnamefont {Linde}}, \
		and\ \bibinfo {author} {\bibfnamefont {D.}~\bibnamefont {Roest}},\ }\href
	{\doibase 10.1103/PhysRevLett.112.011303} {\bibfield  {journal} {\bibinfo
			{journal} {Phys. Rev. Lett.}\ }\textbf {\bibinfo {volume} {112}},\ \bibinfo
		{pages} {011303} (\bibinfo {year} {2014})},\ \Eprint
	{http://arxiv.org/abs/1310.3950} {arXiv:1310.3950 [hep-th]} \BibitemShut
	{NoStop}%
	%%CITATION = ARXIV:1310.3950;%%
	\bibitem [{\citenamefont {Kallosh}\ \emph {et~al.}(2013)\citenamefont
		{Kallosh}, \citenamefont {Linde},\ and\ \citenamefont
		{Roest}}]{Kallosh:2013yoa}%
	\BibitemOpen
	\bibfield  {author} {\bibinfo {author} {\bibfnamefont {R.}~\bibnamefont
			{Kallosh}}, \bibinfo {author} {\bibfnamefont {A.}~\bibnamefont {Linde}}, \
		and\ \bibinfo {author} {\bibfnamefont {D.}~\bibnamefont {Roest}},\ }\href
	{\doibase 10.1007/JHEP11(2013)198} {\bibfield  {journal} {\bibinfo  {journal}
			{JHEP}\ }\textbf {\bibinfo {volume} {11}},\ \bibinfo {pages} {198} (\bibinfo
		{year} {2013})},\ \Eprint {http://arxiv.org/abs/1311.0472} {arXiv:1311.0472
		[hep-th]} \BibitemShut {NoStop}%
	%%CITATION = ARXIV:1311.0472;%%
	\bibitem [{\citenamefont {Podolsky}\ \emph {et~al.}(2006)\citenamefont
		{Podolsky}, \citenamefont {Felder}, \citenamefont {Kofman},\ and\
		\citenamefont {Peloso}}]{Podolsky:2005bw}%
	\BibitemOpen
	\bibfield  {author} {\bibinfo {author} {\bibfnamefont {D.~I.}\ \bibnamefont
			{Podolsky}}, \bibinfo {author} {\bibfnamefont {G.~N.}\ \bibnamefont
			{Felder}}, \bibinfo {author} {\bibfnamefont {L.}~\bibnamefont {Kofman}}, \
		and\ \bibinfo {author} {\bibfnamefont {M.}~\bibnamefont {Peloso}},\ }\href
	{\doibase 10.1103/PhysRevD.73.023501} {\bibfield  {journal} {\bibinfo
			{journal} {Phys. Rev.}\ }\textbf {\bibinfo {volume} {D73}},\ \bibinfo {pages}
		{023501} (\bibinfo {year} {2006})},\ \Eprint
	{http://arxiv.org/abs/hep-ph/0507096} {arXiv:hep-ph/0507096 [hep-ph]}
	\BibitemShut {NoStop}%
	%%CITATION = HEP-PH/0507096;%%
	\bibitem [{\citenamefont {Amendola}\ \emph {et~al.}(2013)\citenamefont
		{Amendola} \emph {et~al.}}]{Amendola:2012ys}%
	\BibitemOpen
	\bibfield  {author} {\bibinfo {author} {\bibfnamefont {L.}~\bibnamefont
			{Amendola}} \emph {et~al.} (\bibinfo {collaboration} {Euclid Theory Working
			Group}),\ }\href {\doibase 10.12942/lrr-2013-6} {\bibfield  {journal}
		{\bibinfo  {journal} {Living Rev. Rel.}\ }\textbf {\bibinfo {volume} {16}},\
		\bibinfo {pages} {6} (\bibinfo {year} {2013})},\ \Eprint
	{http://arxiv.org/abs/1206.1225} {arXiv:1206.1225 [astro-ph.CO]} \BibitemShut
	{NoStop}%
	%%CITATION = ARXIV:1206.1225;%%
	\bibitem [{\citenamefont {Andre}\ \emph {et~al.}(2013)\citenamefont {Andre}
		\emph {et~al.}}]{Andre:2013afa}%
	\BibitemOpen
	\bibfield  {author} {\bibinfo {author} {\bibfnamefont {P.}~\bibnamefont
			{Andre}} \emph {et~al.} (\bibinfo {collaboration} {PRISM}),\ }\href@noop {}
	{\  (\bibinfo {year} {2013})},\ \Eprint {http://arxiv.org/abs/1306.2259}
	{arXiv:1306.2259 [astro-ph.CO]} \BibitemShut {NoStop}%
	%%CITATION = ARXIV:1306.2259;%%
	\bibitem [{\citenamefont {Mao}\ \emph {et~al.}(2008)\citenamefont {Mao},
		\citenamefont {Tegmark}, \citenamefont {McQuinn}, \citenamefont
		{Zaldarriaga},\ and\ \citenamefont {Zahn}}]{Mao:2008ug}%
	\BibitemOpen
	\bibfield  {author} {\bibinfo {author} {\bibfnamefont {Y.}~\bibnamefont
			{Mao}}, \bibinfo {author} {\bibfnamefont {M.}~\bibnamefont {Tegmark}},
		\bibinfo {author} {\bibfnamefont {M.}~\bibnamefont {McQuinn}}, \bibinfo
		{author} {\bibfnamefont {M.}~\bibnamefont {Zaldarriaga}}, \ and\ \bibinfo
		{author} {\bibfnamefont {O.}~\bibnamefont {Zahn}},\ }\href {\doibase
		10.1103/PhysRevD.78.023529} {\bibfield  {journal} {\bibinfo  {journal} {Phys.
				Rev.}\ }\textbf {\bibinfo {volume} {D78}},\ \bibinfo {pages} {023529}
		(\bibinfo {year} {2008})},\ \Eprint {http://arxiv.org/abs/0802.1710}
	{arXiv:0802.1710 [astro-ph]} \BibitemShut {NoStop}%
	%%CITATION = ARXIV:0802.1710;%%
	\bibitem [{\citenamefont {Finelli}\ \emph {et~al.}(2018)\citenamefont {Finelli}
		\emph {et~al.}}]{Finelli:2016cyd}%
	\BibitemOpen
	\bibfield  {author} {\bibinfo {author} {\bibfnamefont {F.}~\bibnamefont
			{Finelli}} \emph {et~al.} (\bibinfo {collaboration} {CORE}),\ }\href
	{\doibase 10.1088/1475-7516/2018/04/016} {\bibfield  {journal} {\bibinfo
			{journal} {JCAP}\ }\textbf {\bibinfo {volume} {1804}},\ \bibinfo {pages}
		{016} (\bibinfo {year} {2018})},\ \Eprint {http://arxiv.org/abs/1612.08270}
	{arXiv:1612.08270 [astro-ph.CO]} \BibitemShut {NoStop}%
	%%CITATION = ARXIV:1612.08270;%%
	\bibitem [{\citenamefont {Boyle}\ and\ \citenamefont
		{Buonanno}(2008)}]{Boyle:2007zx}%
	\BibitemOpen
	\bibfield  {author} {\bibinfo {author} {\bibfnamefont {L.~A.}\ \bibnamefont
			{Boyle}}\ and\ \bibinfo {author} {\bibfnamefont {A.}~\bibnamefont
			{Buonanno}},\ }\href {\doibase 10.1103/PhysRevD.78.043531} {\bibfield
		{journal} {\bibinfo  {journal} {Phys. Rev.}\ }\textbf {\bibinfo {volume}
			{D78}},\ \bibinfo {pages} {043531} (\bibinfo {year} {2008})},\ \Eprint
	{http://arxiv.org/abs/0708.2279} {arXiv:0708.2279 [astro-ph]} \BibitemShut
	{NoStop}%
	%%CITATION = ARXIV:0708.2279;%%
	\bibitem [{\citenamefont {Koh}\ \emph {et~al.}(2018)\citenamefont {Koh},
		\citenamefont {Lee},\ and\ \citenamefont {Tumurtushaa}}]{Koh:2018qcy}%
	\BibitemOpen
	\bibfield  {author} {\bibinfo {author} {\bibfnamefont {S.}~\bibnamefont
			{Koh}}, \bibinfo {author} {\bibfnamefont {B.-H.}\ \bibnamefont {Lee}}, \ and\
		\bibinfo {author} {\bibfnamefont {G.}~\bibnamefont {Tumurtushaa}},\ }\href
	{\doibase 10.1103/PhysRevD.98.103511} {\bibfield  {journal} {\bibinfo
			{journal} {Phys. Rev.}\ }\textbf {\bibinfo {volume} {D98}},\ \bibinfo {pages}
		{103511} (\bibinfo {year} {2018})},\ \Eprint
	{http://arxiv.org/abs/1807.04424} {arXiv:1807.04424 [astro-ph.CO]}
	\BibitemShut {NoStop}%
	%%CITATION = ARXIV:1807.04424;%%
\end{thebibliography}
\end{document}